\documentclass[12pt]{iopart}

\usepackage{graphicx}% Include figure files
\usepackage{dcolumn}% Align table columns on decimal point
\usepackage{bm}% bold math
\usepackage{epsfig}
\begin{document}

\title[Critical manifolds of inhomogeneous bow-tie and checkerboard lattices]{The critical manifolds of inhomogeneous bond
percolation on bow-tie and checkerboard lattices}
\author{Robert M. Ziff}
%\email{rziff@umich.edu}
\address{Michigan Center for Theoretical Physics and Department of Chemical Engineering, University of Michigan, Ann Arbor, Michigan 48109-2136, USA}
\author{Christian R. Scullard}
%\email{scullard1@llnl.gov}
\address{Lawrence Livermore National Laboratory, Livermore, California 94550, USA}
%\affiliation{Department of Applied Mathematics and Statistics, Johns Hopkins University, Baltimore, Maryland, 21218, USA}
\author{John C. Wierman}
\address{Department of Applied Mathematics and Statistics, Johns Hopkins University, Baltimore, Maryland, 21218, USA}

\author{Matthew R. A. Sedlock}
\address{Department of Mathematics and Statistics, Swarthmore College, Swarthmore, PA 19081, USA}
\date{today}

\begin{abstract}
We give a conditional derivation of the inhomogeneous critical percolation manifold of the bow-tie lattice with five different probabilities, a problem that does not appear at first to fall into any known solvable class. Although our argument is mathematically rigorous only on a region of the manifold, we conjecture that the formula is correct over its entire domain, and we provide a non-rigorous argument for this that employs the negative probability regime of the triangular lattice critical surface. We discuss how the rigorous portion of our result substantially broadens the range of lattices in the solvable class to include certain inhomogeneous and asymmetric bow-tie lattices, and that, if it could be put on a firm foundation, the negative probability portion of our method would extend this class to many further systems, including F.Y. Wu's checkerboard formula for the square lattice. We conclude by showing that this latter problem can in fact be proved using a recent result of Grimmett and Manolescu for isoradial graphs, lending strong evidence in favour of our other conjectured results.
\end{abstract}

\maketitle
%\pacs{Ak 64.60}% PACS, the Physics and Astronomy
%\maketitle
\section{Introduction}
Finding critical probabilities and critical manifolds (for inhomogeneous systems)
is a longstanding problem in understanding percolation \cite{SykesEssam,Wu79,Wierman84,StaufferAharony94}, and continues
to be the subject of much study today \cite{ScullardZiff08,FengDengBlote08,Scullard11,WiermanZiff11,GrimmettManolescu11,MinnhagenBaek10,Wu10,DingFuGuoWu10}. Given any lattice, in a homogeneous percolation model we declare each bond to be open with probability $p$ and closed with complementary probability $1-p$. The size of connected open clusters grows with $p$, and there is a lattice-dependent critical probability, $p_c$, above which there is an infinite cluster. The problem can be generalized in several ways, one being the inhomogeneous model that assigns different probabilities, $(p_1,p_2,...,p_n)$, to different bonds. The problem now is, with $n-1$ of these probabilities set to arbitrary values, to find the critical value of the remaining probability. The solution may be written as a function
\begin{equation}
f(p_1,p_2,...,p_n)=0
\end{equation}
where $f$ is the critical surface, or critical manifold, of the problem.

\begin{figure} %[htbp]
\centering
%\vspace{0.2in}
\includegraphics{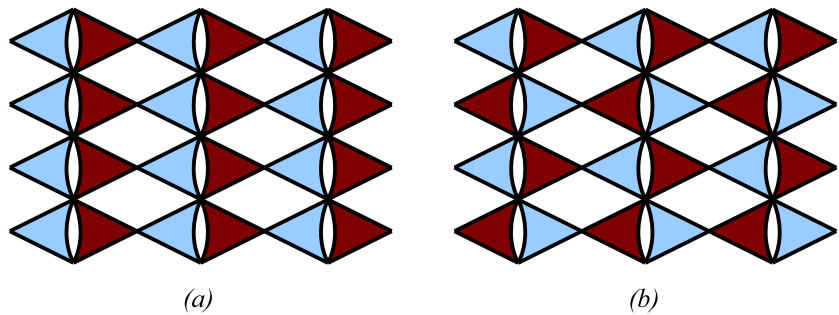}
\caption{A self-dual 3-uniform hypergraph. The colored regions can represent any network of bonds. If they are simple triangles, then we have the split-bond bow-tie lattice \cite{Wierman84}. Alternating (red and blue) generators can be used in the configurations a) and b).}
\label{fig:3uniform}
\end{figure}

Recently it has been shown that the critical manifold for a large class of graphs formed using 3-uniform hypergraphs can be found exactly \cite{WiermanZiff11,Ziff06,ScullardZiff06,ChayesLei06,BollobasRiordan11}. An example of such a hypergraph is shown in Figure \ref{fig:3uniform}, where the colored regions, called hyperedges, are not necessarily simple triangles but can represent any network of bonds (including correlated bonds as well as internal sites) contained between three boundary vertices. However, we assume for the moment that all hyperedges are identical, with identical assignments of probabilities (i.e., ignore the colors for now). The dual of a planar 3-uniform hypergraph is also a 3-uniform hypergraph, and, if the hypergraph is self-dual, the critical point can be located using a simple condition \cite{Ziff06}: the probability that all three vertices of a
hyperedge connect is equal to the probability that none connect, or,
\begin{equation}
 \hbox{Prob(all) = Prob(none)} .
\label{eq:allequalsnone}
\end{equation}
Condition (\ref{eq:allequalsnone}) ensures that the triangle-triangle transformation, i.e., for bond percolation, replacing the hyperedges, now realized as collections of bonds, on the original lattice by their duals and replacing the probabilities with their complements (e.g., $p \rightarrow 1-p$), leaves the connectivity of the boundary vertices invariant. That this implies criticality for a broad class of lattices is by now well established rigorously \cite{WiermanZiff11,BollobasRiordan11}. In addition, as long as (\ref{eq:allequalsnone}) is satisfied on all our bond-realized hyperedges, we may replace any hyperedge by its dual with complemented probabilities, and the system will remain at criticality.

\begin{figure} %[htbp]
\centering
%\vspace{0.2in}
\includegraphics{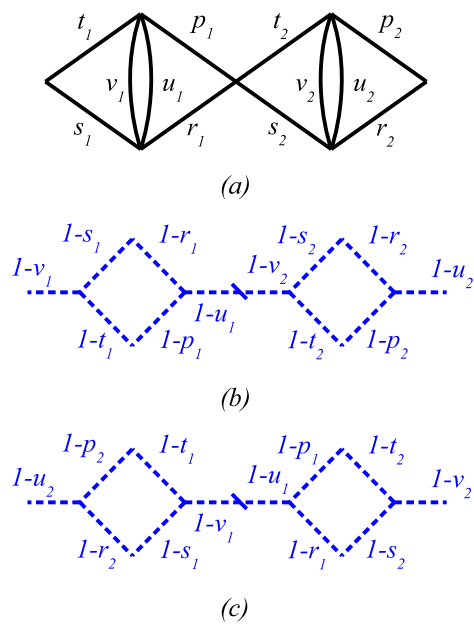}
\caption{a) An assignment of probabilities over four triangles of the split-bond bow-tie lattice; b) the system resulting from the triangle-triangle (or star-triangle \cite{SykesEssam}) transformation; c) the dual system (shifted), which is inequivalent to b). These figures are repeated periodically to form their respective lattices.}
\label{fig:BTtrans}
\end{figure}

In this paper, we will drop the requirement that each hyperedge be identical, a generalization that is also discussed in \cite{BollobasRiordan11}. Then the applicability of (\ref{eq:allequalsnone}) for critical manifolds depends not only on the lattice, but also on the way in which probabilities are assigned to the bonds. Consider the situation in Figure \ref{fig:BTtrans}, in which different probabilities are assigned over four triangles. Here, employing the triangle-triangle transformation by simply replacing each triangle by its dual with complementary probabilities produces the system in Figure \ref{fig:BTtrans}b. However, this is not the dual process (Figure \ref{fig:BTtrans}c) --- the underlying lattice is indeed the graph-theoretic dual, but the probabilities are scrambled. If the triangle-triangle transformation does not produce the correct dual model, equation (\ref{eq:allequalsnone}) cannot give the critical manifold.

By considering this generalization, we will obtain critical manifolds, both rigorous and non-, on lattices that can be cast in the form of those shown in Figure \ref{fig:3uniform}. Although many of these have unit cells contained between four boundary vertices rather than the usual three, this method is not applicable to {\it general} lattices of this type, such as the kagome lattice.

\section{The checkerboard model critical manifold} 
\begin{figure} %[htbp]
\centering
%\vspace{0.2in}
\includegraphics{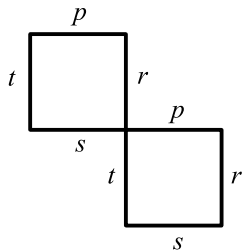}
\caption{The bonds on the inhomogeneous checkerboard lattice}
\label{fig:checkerboard}
\end{figure}

\nonumber
One system with a conjectured exact solution that apparently cannot
be verified directly by (\ref{eq:allequalsnone}) is the 
``checkerboard" lattice, which is just the square lattice but with the probability assignments shown in Fig.\ \ref{fig:checkerboard}. 
In 1979, Wu \cite{Wu79} proposed that the critical manifold for this system is given by $C(p, r, s, t) = 0$, where
\begin{eqnarray}
C(p, r, s, t) &\equiv& 1 - p r - p s - r s - p t - r t \cr
&-& s t + p r s + p r t + r s t + p s t . \label{eq:checkerboard}
\end{eqnarray}
This formula was also found by some of the present authors \cite{ScullardZiff08} by considering 
the general mathematical form first-order in all the probabilities and which reduces to known 
exact results (square lattice, honeycomb lattice, etc.) in the appropriate limits of 
$p$, $r$, $s$ and $t$. Computer simulations \cite{ScullardZiff08} verified this
result to more than six-digit accuracy at several points.
Equation (\ref{eq:checkerboard}) also follows from more general arguments in the 
Potts model \cite{Wu79,MaillardRammal85}, however, some doubt was raised that the $q-$state manifold is actually correct \cite{Enting}.
Our goal here is to derive (\ref{eq:checkerboard}) directly from percolation duality
arguments. From a mathematically rigorous standpoint, we will only be partially successful in this as our argument relies crucially on a step that makes use of the negative probability region of the triangular lattice critical manifold. Nevertheless, we will conjecture that this regime is meaningful and that critical manifolds derived in this way are correct. We will also show that our conjecture implies an even wider class of inhomogeneous solutions than (\ref{eq:checkerboard}), leading to an array of new lattices with critical thresholds that can apparently (and unexpectedly) be found exactly.

\begin{figure} %[htbp]
\centering
%\vspace{0.2in}
\includegraphics{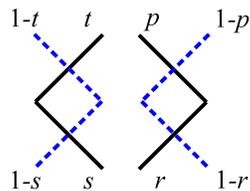}
\caption{The decomposition of the checkerboard into triangles. The dual triangles are in dotted blue.}
\label{fig:twotriangles}
\end{figure}

One special case in which (\ref{eq:checkerboard}) is easy to verify is where
\begin{equation}
p + r = 1, \qquad s + t = 1  \ . \label{eq:exact}
\end{equation}
Here, we present a simple example of our general approach by breaking the lattice into two types of hyperedges, as shown in Fig.\ \ref{fig:twotriangles} with the dual hyperedges shown in blue (dotted), and the triangle-triangle transformation yields the dual rotated 180 degrees. Equation \ (\ref{eq:allequalsnone}) can be applied separately to each of the two hyperedges, and gives (\ref{eq:exact}). However, a direct examination of the general checkerboard lattice does not reveal a simple way to relate the four bond probabilities.
\begin{figure} %[htbp]
\centering
%\vspace{0.2in}
\includegraphics{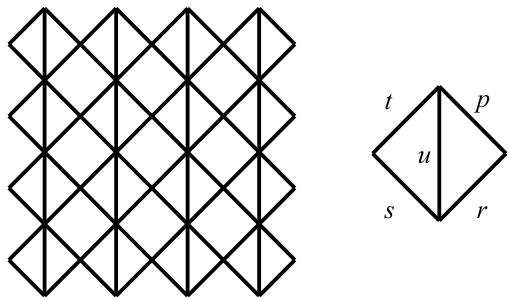}
\caption{The bow-tie lattice with the assignment of probabilities on the right. The critical manifold for this system is given by equation (\ref{eq:bowtie}).}
\label{fig:bowtie}
\end{figure}

\section{A duality-based derivation}
The key to applying duality for the checkerboard is to put diagonals into alternate squares and split them into two, forming 
the split-bond bow-tie lattice \cite{Wierman84} (see Fig.\ \ref{fig:3uniform}). In fact, in \cite{ScullardZiff08} the threshold for the inhomogeneous bow-tie lattice shown in Figure \ref{fig:bowtie}
was also conjectured by assuming a consistent multi-linear form, with the result $B(p,r,s,t,u)=0$, where
\begin{eqnarray}
B(p, r, s, t, u) &=& 1 - u - p r - p s - r s - p t - r t \cr 
&-& s t + p r s + p r t + p s t + r s t + p r u \cr
&+& s t u - p r s t u \label{eq:bowtie} \\
 &=& C(p, r, s, t) - u (1 - p r - s t + p r s t)
\end{eqnarray}
We can establish this result for a subset of the parameter space by considering the construction shown in Figure \ref{fig:TTtrans}, in which the right- and left-pointing triangles on the split-bond bow-tie lattice are given the probabilities $(p,r,u_1)$ and $(s,t,u_2)$, respectively. Separately imposing the criterion (\ref{eq:allequalsnone}) on the different triangles results in a critical system and gives the separate conditions,
\begin{equation}
 p r u_1 - p - r - u_1 + 1 = 0
\end{equation}
and
\begin{equation}
 s t u_2 - s - t - u_2 + 1 = 0 ,
\end{equation}
which, provided $p+r \le 1$ and $s+t \le 1$ to ensure that the probabilities $u_1, u_2 \ge 0$, can be written,
\begin{equation}
 u_1=\frac{1-p-r}{1-pr} \label{eq:u1}
\end{equation}
and
\begin{equation}
 u_2=\frac{1-s-t}{1-st} . \label{eq:u2}
\end{equation}
This appears somewhat trivial, since we have just set the different triangles independently to their critical values, as before. However, the bonds $u_1$ and $u_2$ can be combined into a single effective bond of probability $u$, a trick originally employed by one of the present authors \cite{Wierman84} to find the homogeneous bow-tie bond threshold (this argument was recently generalized to the $q-$state Potts model in \cite{Ding12}). In order to cross this bond, it is necessary that at least one of $u_1$ and $u_2$ be open, i.e.,
\begin{equation}
u=1-(1-u_1)(1-u_2)=u_1+u_2-u_1 u_2 . \label{eq:u}
\end{equation}
Using (\ref{eq:u1}) and (\ref{eq:u2}) in (\ref{eq:u}) immediately gives the inhomogeneous bow-tie result (\ref{eq:bowtie}), and the checkerboard when we set $u=0$. However, this last step is more subtle than it appears, as we discuss in the next section.

\section{Balancing super-critical and sub-critical triangles}
Everything up to equation (\ref{eq:u}) has been grounded in rigorous theory. In particular, in the terminology of \cite{BollobasRiordan11}, we have a self-dual hyperlattice percolation model, and so by their Theorem 2.1, the system is critical. We may thus consider equation (\ref{eq:bowtie}) to be firmly established, provided we require $p+r \le 1$ and $s+t \le 1$. However, there is nothing in the formula (\ref{eq:bowtie}) that suggests such stringent inequalities are necessary and it is tempting to speculate that (\ref{eq:bowtie}) actually holds more generally. In fact, setting $u=0$ does not even give us the checkerboard unless we allow either $p+r>1$ or $s+t>1$; according to (\ref{eq:u}), setting $u=0$ means $u_1=u_2=0$, and thus $p+r=s+t=1$. 

\begin{figure} %[htbp]
\centering
%\vspace{0.2in}
\includegraphics{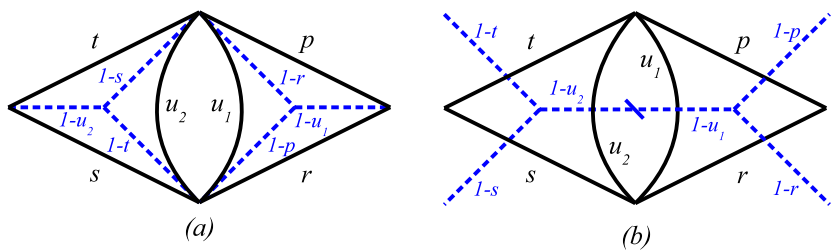}
\caption{a) The triangle-triangle transformation on the bow-tie lattice with a split bond; b) the dual transformation. The triangle-triangle transformation yields the dual rotated 180 degrees.}
\label{fig:TTtrans}
\end{figure}

On the other hand, the critical manifold for the checkerboard, (\ref{eq:checkerboard}), if it is correct, allows one to select three of the probabilities, say $p$, $r$ and $s$, arbitrarily, and then provides the value of $t$ to make the system critical. One might choose, for example, $p$ and $r$ to make the right-pointing triangle super-critical, i.e. $p+r>1$, but the result is that $u_1$ becomes negative. In fact, it might take any value in $(-\infty,0]$, even though both $p$ and $r \in [0,1]$, and so in this regime, $u_1$ is not a probability but a parameter, which we call the criticality parameter, that measures the degree of super-criticality in the triangle. Correspondingly, in equation (\ref{eq:u}), choosing $u_1$ or $u_2$ (but not both) negative is formally allowed so long as the resulting $u$ from (\ref{eq:u}) is in $[0,1]$. Suppose we choose $u_1$ to have some value in $(-\infty,0]$, then it is easy to see that for any $u \in [0,1]$ there is a $u_2 \ge u$ also in $[0,1]$ that satisfies (\ref{eq:u}). As such, for $u_1<0$, we can interpret equation (\ref{eq:u}) as describing a bond with probability $u_2$ that is attached to a sink, making the actual traversal probability $u$. This effect is described by the parameter $u_1 \in (-\infty,0]$, which we call the sink parameter. Now, if we set $p$ and $r$ to super-critical values, then in order for the whole system to be critical, it is necessary that the left-pointing triangle be sub-critical. Clearly, it can be made so by sufficiently lowering the probability of the $u_2$ bond with some sink. The conjecture that lies at the heart of our extension of (\ref{eq:bowtie}) is that criticality is achieved by setting the sink parameter equal to the criticality parameter, and in this way balancing the super-criticality on the right with sub-criticality on the left. Now we may set $u=0$ without needing $u_1=u_2=0$ because $u_1<0$. In this way, we speculatively extend the validity of the inhomogeneous bow-tie manifold (\ref{eq:bowtie}) to cases in which either $p+r>1$ and $s+t<1$, or $p+r<1$ and $s+t>1$, and recover the checkerboard as the special case $u=0$. Note that $p+r$ and $s+t$ still cannot both be greater than one because then the system is super-critical regardless of the value of $u$. We note here that we might also have formulated this procedure in the dual process (the dotted lines in Figure \ref{fig:TTtrans}). In this case, rather than having a negative probability, one of the probabilities, say $1-u_1$, would be greater than 1. Now the two sides are connected through a bond, composed of two edges in series, with probability $(1-u_1)(1-u_2)$ and this effective probability is in the appropriate range, $[0,1]$.

One somewhat puzzling property of the checkerboard manifold (\ref{eq:checkerboard}) is its $S_4$ symmetry \cite{MaillardRammal85} --- it is invariant under the interchange of any two probabilities. Most of these result from ordinary rotation, translation, and reflection symmetries, but the fact that one can switch two adjacent bonds without also switching the other two is rather surprising. However, our derivation makes it clear how this symmetry arises; in Figure \ref{fig:TTtrans}a, it makes no difference to the result if we flip, say, the right triangle about a horizontal line through its center without flipping the other one.

The checkerboard manifold for percolation has previously been tested numerically \cite{ScullardZiff08}. Setting $p=73/90$, and $s=t=r$ in (\ref{eq:checkerboard}) leads to the prediction $r_c=0.4$. Using the hull-gradient method, it was found that $r_c=0.400\ 000\ 04(10)$, completely consistent with the formula (the number in brackets is the standard deviation in the last digits). As already mentioned, for the checkerboard formula to make the correct prediction, either $u_1$ or $u_2$ must be negative, and in this case $u_1=-5/16$. Thus, the numerical result can be seen as strong support for the above negative ``probability'' argument. The inhomogeneous bow-tie formula (\ref{eq:bowtie}) was checked using the configuration $u=r=s=p$ and $t=1/2$, which predicts $p_c=0.38196601... $\ . Numerically, $p_c=0.3819654(5)$, placing our prediction easily inside two standard deviations. However, for these parameters, both $u_1$ and $u_2$ are positive, and thus we are in the completely rigorous regime of equation (\ref{eq:bowtie}).
\begin{figure}%[htbp]
\centering
%\vspace{0.2in}
\includegraphics{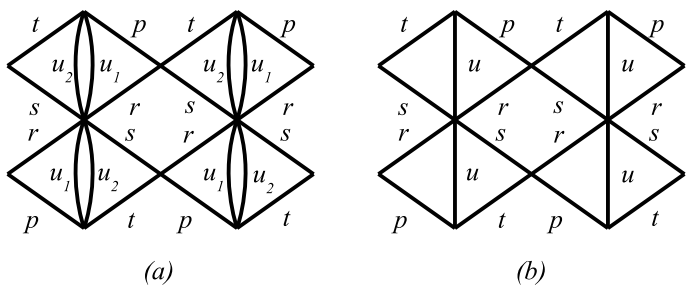}
\caption{a) An assignment of probabilities on the split-bond bow-tie lattice which is self-dual under the triangle-triangle transformation; b) the corresponding bow-tie lattice with the bond merged. This system differs from that in Figure \ref{fig:bowtie} in that every second row is rotated 180 degrees.}
\label{fig:bowtierot}
\end{figure}
\section{Extension to asymmetric bow-tie lattices}
\begin{figure}%[htbp]
\centering
%\vspace{0.2in}
\includegraphics{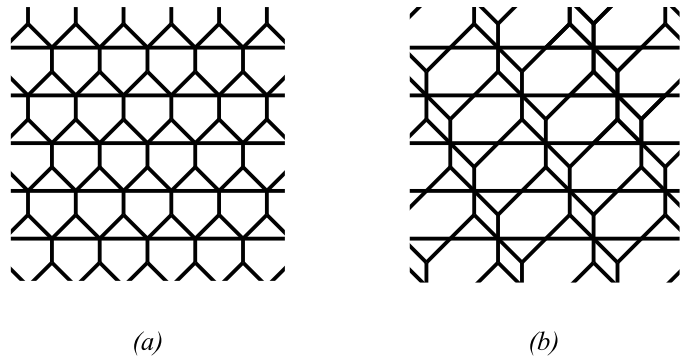}
\caption{a) The self-dual martini-B lattice, which results from setting $r=0$ in Figure \ref{fig:bowtie}b; b) the self-dual lattice that results from setting $r=0$ in Figure \ref{fig:bowtierot}b. These two lattices have identical critical manifolds.}
\label{fig:Blattices}
\end{figure}

Our construction has implications for systems other than the inhomogeneous bow-tie and checkerboard lattices. For example, consider Figure \ref{fig:bowtierot}, in which triangles in each row are rotated $180$ degrees relative to the row above, a particular realization of Figure \ref{fig:3uniform}b. The triangle-triangle transformation again yields the dual, and thus the solution will also be given by (\ref{eq:bowtie}). It is not obvious that Figures \ref{fig:bowtie} and \ref{fig:bowtierot} should have the same solution. For example, setting $r=0$ in Figure \ref{fig:bowtie} results in the self-dual martini-B lattice \cite{ZiffScullard06,Scullard06,Wu06} (Figure \ref{fig:Blattices}a), whereas setting $r=0$ in Figure \ref{fig:bowtierot}b also gives a self-dual lattice (Figure \ref{fig:Blattices}b) but it bears only a slight resemblance to the martini-B lattice. Nevertheless, these lattices have identical inhomogeneous critical manifolds and for the homogenous case, both have $p_c = 1/2$.

\begin{figure} [htbp]
\centering
%\vspace{0.2in}
\includegraphics{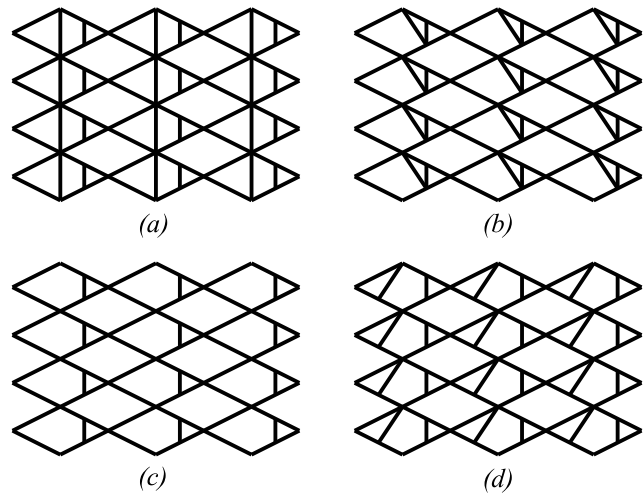}
\caption{Asymmetric bow-tie lattices, for which exact bond thresholds can now be found: c) $1 - p - p^2 - 4 p^3 - 2 p^4 + 15 p^5 - 10 p^6 + p^8=0$, $p_c=0.481216...$; b) $1 - 2 p^2 - 7 p^3 + 25 p^5 - 27 p^6 + 10 p^7 - p^8=0$, $p_c=0.516867...$; a) $1 - p^2 - 5 p^3 - 3 p^4 + 14 p^5 - 8 p^6 + p^7=0$, $p_c=0.575716...$ ; d) $1 - p^2 - 4 p^3 - 6 p^4 + 9 p^5 + 14 p^6 - 22 p^7 + 9 p^8 - p^9=0$, $p_c=0.566302...$ .}
\label{fig:ASbowtie}
\end{figure}

In addition to this freedom to alternate the triangles in neighboring rows, it is also not necessary that the left- and right-pointing triangles contain the same generators. Figures \ref{fig:3uniform}a and b give possible ways to arrange alternating triangular generators. Consider Figure \ref{fig:ASbowtie}a, in which we use the martini-A lattice generator on the right and the triangular generator on the left. More specifically, we would put a generator for the martini-A lattice with a bond across the bottom in the right triangle, a simple triangle in the left, and combine the central parallel bonds by the above procedure. The result is a kind of asymmetric bow-tie lattice, which, at first glance, does not obviously fall into the solvable 3-uniform class. However, the present argument shows that it does, and in this case we do not need to make use of negative probabilities, so its exact bond threshold is given rigorously by the solution of $1 - p - p^2 - 4 p^3 - 2 p^4 + 15 p^5 - 10 p^6 + p^8=0$, $p_c=0.481216...$ . Of course, there is an endless variety of these kinds of lattices, and we give a small sampling in Figure \ref{fig:ASbowtie} with their bond thresholds given in the caption. While the threshold in Figure \ref{fig:ASbowtie}a is rigorous, the rest rely on the negative probability conjecture. Thus, even though the rigorous part of our argument expands the class of solvable lattices, from a mathematical perspective it is only a part of what would be obtained by a proof that the negative probability regime of the triangular critical manifold has the meaning and utility we have conjectured.

\section{Relation to isoradial percolation}
\begin{figure} %[htbp]
\centering
%\vspace{0.2in}
\includegraphics{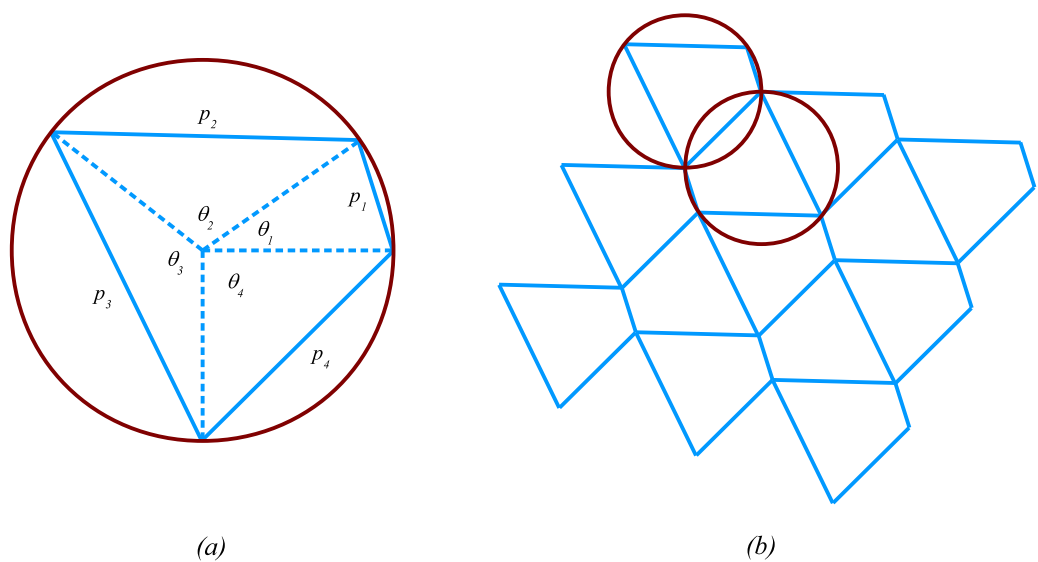}
\caption{a) Isoradial construction for a general inhomogeneous checkerboard lattice, with the angles $\theta_i$ related to the probabilities by (\ref{eq:tanformula}); b) tiling of the plane with quadrilaterals, resulting in an isoradial checkerboard. A circle of fixed radius can be placed with its center inside each face such that the vertices of the face lie on the circle's circumference.}
\label{fig:isoradial}
\end{figure}

Recently, Grimmett and Manolescu \cite{GrimmettManolescu12} proved a conjecture of Kenyon \cite{Kenyon} on the criticality of a class of graphs that can be embedded in the plane such that all polygons of bonds fall on circles of identical radii with centers enclosed within the polygons. This embedding is equivalent to covering the plane with a tiling of rhombi, such that the edges connecting opposite vertices form the lattice and dual-lattice bonds.  According to the formula of Kenyon, criticality corresponds to assigning a probability $p_i$ to a bond that subtends an angle $\theta_i$ from the center of the circles (see Figure \ref{fig:isoradial}a) by the formula
\begin{equation}
\frac{p_i}{1-p_i} = \frac{\sin([\pi - \theta_i]/3)}{\sin(\theta_i/3)}
\label{eq:sinformula}
\end{equation} or 
\begin{equation}
\tan(\theta_i/3) = \sqrt 3 \ \frac{1 - p_i}{1 + p_i}.
\label{eq:tanformula}
\end{equation}
It turns out that this criticality condition can be used to prove Wu's checkerboard criticality condition (\ref{eq:checkerboard}).  In Fig.\ \ref{fig:isoradial}b we show a tiling of the plane by a general quadrilateral that falls on an isoradial circle.  Evidently, we can tile the plane with these by alternately flipping the quadrilaterals in a pattern that exactly emulates the repetition of probabilities on the checkerboard lattice.  Say that the four angles subtending the four arbitrary bonds are $\theta_1$, $\theta_1$, $\theta_3$ and $\theta_4$, such that $\theta_1 + \theta_1 + \theta_3 + \theta_4 = 2 \pi$.  We use the identity
\begin{equation}
\tan(a+b+c+d) = \frac{u_a + u_b + u_c + u_d- u_{abc}  - u_{abd} - u_{acd} - u_{bcd}}{1 - u_{ab} - u_{ac} - u_{ad} - u_{bc} - u_{bd} - u_{cd} + u_{abcd}}
%\frac{ \tan a + \tan b + \tan c + \tan d}{1 - \tan a \tan b - \tan a \tan c} \cdots \cr 
%&\cdots&\frac{ - \tan a \tan b \tan c -\tan a \tan b \tan d - \tan a \tan c \tan d }{- \tan a \tan d - \tan b \tan c - \tan b \tan d - \tan c \tan d } \cdots \cr
%&\cdots&\frac{- \tan b \tan c \tan d}{+ \tan a \tan b \tan c \tan d}
\label{eq:tanabcd}
\end{equation}
where $u_a \equiv \tan a$, $u_{ab} \equiv \tan a \tan b$, etc., with $a = \theta_1/3$, $b = \theta_2/3$, $c = \theta_3/3$, and $d = \theta_4/3$, and use (\ref{eq:tanformula}) to relate $\tan \theta_i/3$ to $p_i$ (with $p_1 = p$, $p_2 = r$, $p_3 = s$, and $p_4 = t$).  Setting the result of (\ref{eq:tanabcd}) to $\tan(2 \pi / 3) = - \sqrt{3}$, and after some computer algebra, we indeed find that Wu's formula (\ref{eq:checkerboard}) results.  Thus, the $S_4$ symmetry inherent in Wu's formula follows from the isoradial criticality result (\ref{eq:sinformula}).

On the other hand, it does not seem possible to put the bow-tie lattice in an isoradial embedding, and consequently we cannot use the isoradial result to prove that formula.  However, because the bow-tie formula gives the checkerboard formula in the limit that $u=0$, and also because the general bow-tie formula is definitely valid for cases in which $u_1 > 0$ and $u_2>0$, it seems highly likely that the bow-tie formula (\ref{eq:bowtie}) is valid for all values of its parameters.

\section{Conclusion}
We have generalized the criticality condition (\ref{eq:allequalsnone}) to systems of inhomogeneous 3-hypergraphs, proposing several new lattices with exact critical manifolds.  Applying this idea to the bow-tie lattice, we are able to obtain the previously conjectured manifolds for the inhomogeneous bow-tie and checkerboard lattices, although for the latter and some cases of the former, we must introduce intermediate bonds with negative probability.  The meaning of such bonds is not entirely clear, although for the checkerboard case at least this approach is confirmed by the isoradial construction.  In the final results (\ref{eq:checkerboard}) and (\ref{eq:bowtie}), of course, all bonds have positive probability in $[0,1]$.

The authors thank Geoffrey Grimmett for suggesting that the checkerboard result can be proven by using the isoradial mapping. John Wierman and Matthew Sedlock acknowledge the support of the Acheson J. Duncan Fund for Research in Statistics. This work was partially performed under the auspices of the U.S. Department of Energy by Lawrence Livermore National Laboratory under Contract DE-AC52-07NA27344.

\section*{References}
\bibliography{ziffscullard11}% Produces the bibliography via BibTeX.

\end{document}